\documentstyle[12pt]{article}

\begin{document}
\newcommand{\pl}[1]{Phys.\ Lett.\ {\bf #1}\ }
\newcommand{\npb}[1]{Nucl.\ Phys.\ {\bf B#1}\ }
\newcommand{\prd}[1]{Phys.\ Rev.\ {\bf D#1}\ }
\newcommand{\prl}[1]{Phys.\ Rev.\ Lett.\ {\bf #1}\ }

\newcommand{\drawsquare}[2]{\hbox{%
\rule{#2pt}{#1pt}\hskip-#2pt
\rule{#1pt}{#2pt}\hskip-#1pt
\rule[#1pt]{#1pt}{#2pt}}\rule[#1pt]{#2pt}{#2pt}\hskip-#2pt
\rule{#2pt}{#1pt}}

\newcommand{\Yfund}{\raisebox{-.5pt}{\drawsquare{6.5}{0.4}}}
\newcommand{\Ysymm}{\raisebox{-.5pt}{\drawsquare{6.5}{0.4}}\hskip-0.4pt%
        \raisebox{-.5pt}{\drawsquare{6.5}{0.4}}}
\newcommand{\Yasymm}{\raisebox{-3.5pt}{\drawsquare{6.5}{0.4}}\hskip-6.9pt%
        \raisebox{3pt}{\drawsquare{6.5}{0.4}}}

\begin{titlepage}
\begin{center}
{\hbox to\hsize{hep-th/9607210 \hfill  MIT-CTP-2552}}
{\hbox to\hsize{               \hfill  BU/HEP-96-23}}

\bigskip

\bigskip

{\Large \bf  Exact Results and Duality for $SP(2N)$ SUSY Gauge Theories
             with an Antisymmetric Tensor     } \\

\bigskip

\bigskip

{\bf Csaba Cs\'aki and Witold Skiba}\\

\smallskip

{ \small \it Center for Theoretical Physics,

Massachusetts Institute of Technology, Cambridge, MA 02139, USA }

\smallskip

{\tt csaki@mit.edu, skiba@mit.edu}

\bigskip

{\bf Martin Schmaltz}\\

{\small \it Department of Physics, Boston University, Boston, MA 02215, USA }

\smallskip

{\tt schmaltz@abel.bu.edu}

\vspace{1cm}
{\bf Abstract}\\
\end{center}

\bigskip
We study supersymmetric $Sp(2N)$ gauge theories with matter in the
antisymmetric tensor representation and $F$ fundamentals. For $F=6$
we solve the theory exactly in terms of confined degrees of freedom
and a superpotential. By adding mass terms we obtain the theories
with $F<6$ which we find to exhibit a host of interesting non-perturbative
phenomena: quantum deformed moduli spaces with $N$ constraints,
instanton-induced superpotentials and non-equivalent disjoint branches
of moduli spaces. We find a simple dual for $F=8$ and no superpotential.
We show how the $F=4$ and $F=2$ theories can be modified
to break supersymmetry spontaneously and point out that the $Sp(6)$ theory
with $F=6$ may be very interesting for model builders.

\bigskip

\end{titlepage}

\section{Introduction}

There is an increasing number of asymptotically-free $N=1$ supersymmetric
(SUSY) gauge theories that can be analyzed non-perturbatively using
techniques recently discovered by Seiberg~\cite{exact,duality}.
These theories exhibit many
interesting phenomena generated by the non-perturbative dynamics.

The ``classic" example is SUSY QCD: an $SU(N_c)$ gauge theory with $N_f$
fields in fundamental and antifundamental representations. For $N_f < N_c$,
there is a dynamically generated superpotential, which is singular at the
origin~\cite{ADS}. For $N_f=N_c$, the infrared theory confines and
the classical constraint on the meson and baryon fields is modified
quantum mechanically, resulting in chiral symmetry breaking~\cite{exact}.
For $N_f=N_c+1$, the low-energy theory is confining without chiral symmetry
breaking~\cite{exact}. For $N_c+1<N_f<3 N_c$, the theory has a dual description
in terms of an $SU(N_f-N_c)$ gauge theory~\cite{duality}. 
The two descriptions are completely equivalent in the infrared.

A similar analysis can be carried out for other gauge groups. $SO(N)$ and
$Sp(2N)$ theories have been analyzed in Refs.~\cite{duality,IntrSeiberg}
and Refs.~\cite{duality,IntPoul}, respectively. Theories with more complicated
matter have been analyzed as well. Examples include $SU(N)$ theories with an
antisymmetric tensor~\cite{PoppitzTriv,Pouliot} and $SU(N)$, $SO(N)$ and
$Sp(2N)$ theories including matter in the adjoint
representation~\cite{Kutasov,Martin}. There are also similar results
for product groups~\cite{ILS,PST}.

In this paper, we analyze $Sp(2 N)$ gauge theories
containing one antisymmetric tensor in addition to $F$ fields in the 
fundamental representation. We find that $Sp(2 N)$ theories for $F=6$
confine without breaking chiral symmetries and can be described by
an exact superpotential. We identify the low-energy spectrum and derive
the superpotential. Using this superpotential we obtain the
superpotential for $F<6$ by integrating out matter. For $F=4$, we find that
the theory is still confining, but the moduli space is constrained.
One of the classical constraints is modified quantum mechanically,
while the remaining $N-1$ constraints are unmodified. By integrating out
more matter, we find the dynamically generated superpotentials for $F=0,2$.

For $F=8$, we obtain a simple dual for the theory without the addition
of a superpotential. This dual has the same $Sp(2N)$ gauge group and
contains extra gauge singlet meson fields and a superpotential. This 
is the first example in the literature of a simple dual of a theory
containing a two-index tensor and no superpotential. However, it
seems very difficult to extend this duality for $F>8$.
After addition of a superpotential that breaks some of the
global symmetries and simplifies the infrared theory by lifting some of the 
classical flat directions, a dual description can be found 
for arbitrary $F$~\cite{ken}.

These results about $Sp(2N)$ can have several important applications 
to model building. The fact that $Sp(6)$ with $F=6$ has three 
antisymmetric tensors in its low-energy description can be
used for building of composite models. We present a simple toy model
based on $Sp(6)$ that naturally generates a hierarchy for fermion
masses. Our results can also be applied to construct models of 
dynamical supersymmetry breaking. 
We give one example of supersymmetry breaking by an instanton-generated
superpotential and one example with a quantum-deformed moduli space.

This paper is organized as follows: in the next section, we solve
the theories for $F\leq 6$ and present consistency checks on our solutions.
Section 3 is devoted to the theories with $F>6$.
Section 4 contains applications
of our results to dynamical supersymmetry breaking and model building.
Finally, we conclude in Section 5.

\section{Confinement and Exact Superpotentials for $F\leq 6$}

We consider supersymmetric $Sp(2 N)$ gauge theories containing an
antisymmetric tensor $A$, and $F$ fields $Q_i$ in the fundamental
representation. The Witten anomaly requires that $F$ is
even~\cite{Witten}.

The non-anomalous global symmetry of the microscopic theory is
$SU(F) \hspace{-1pt} \times U(1) \times U(1)_R$, with one possible charge
assignment given in the table below.
\begin{equation}
 \begin{array}{c|cccc}
    & Sp(2 N) & SU(F) & U(1) & U(1)_R \\ \hline
  A & \Yasymm   & 1     & -\frac{F}{2}   & 0 \\
 Q_i& \Yfund    & \Yfund& N-1& 1-\frac{4}{F}
 \end{array}
\end{equation}
Note that for $F=4$ one can define an anomaly free R-symmetry under which
all chiral superfields carry charge zero. In analogy with SUSY QCD, we 
expect that this theory might confine and have a quantum deformed moduli
space. Furthermore, we then expect that the theory with $F=6$ confines
without chiral symmetry breaking and that the proper degrees of freedom
in the infrared are given by the set of independent gauge-invariant
operators
\begin{eqnarray}
T_k &=& \frac{1}{4k} {\rm Tr} A^k, \ k=2,3,\ldots,N, \\
M_k &=& Q A^k Q, \ k=0,1,2,\ldots,N-1,
\end{eqnarray}
where all contractions are performed using the $Sp(2N)$ invariant 
antisymmetric tensor $J=i \sigma_2 \otimes 1_{N \times N}$.
The symmetry properties of these fields for $F=6$ 
are given in the table below:
\begin{equation}
 \begin{array}{c|ccc}
         & SU(6)   & U(1) & U(1)_R \\ \hline
  T_k    & 1       & -3k  & 0 \\
  M_k    & \Yasymm & 2 (N-1) -3 k & \frac{2}{3} 
  \end{array}
\end{equation}
It is a very non-trivial consistency check that for $F=6$ the global
anomalies of the ultraviolet theory are matched by these gauge invariants.
The global anomalies in both microscopic and macroscopic descriptions are
\begin{eqnarray}
  SU(6)^3      &=& 2 N \nonumber \\
  SU(6)^2 U(1) &=& 2 N (N-1) \nonumber \\
  SU(6)^2 U(1)_R &=& -\frac{4}{3} N \nonumber \\
  U(1)^3 &=& -27 [ N (2N-1) -1] + 12 N (N-1)^3 \nonumber \\
  U(1)^2 U(1)_R &=& (1-N) (8 N^2 + 10 N +9) \nonumber \\
  U(1)_R^2 U(1) &=& \frac{1}{3}(1-N)(2 N+9) \nonumber \\
  U(1)_R^3 &=& 1-\frac{14}{9} N \nonumber \\
  U(1) &=& 3 (N-1)(2 N-1) \nonumber \\
  U(1)_R &=& 1-6 N
\end{eqnarray}
We believe that, technically, this is the most non-trivial example of
anomaly matching in the literature; it involves the use of
the identities: $\sum_1^n k^2 = \frac{n(n+1)(2n+1)}{6}$ and
$\sum_1^n k^3=\frac{n^2 (n+1)^2}{4}$. 
The fact that one can match the anomalies with this set of operators
strongly suggests that, as expected, the $F=6$ theories are in a phase of
confinement without chiral symmetry breaking, where the low-energy degrees
of freedom are given by the $M$'s and $T$'s above.

Using symmetry arguments and demanding that the equations of motion reproduce
the classical constraints uniquely determine the non-perturbative
superpotential in the gauge-invariant fields.
Since the number of terms in the superpotential
grows rapidly with $N$ we only list the superpotentials for $N=2,3,4$:
\begin{eqnarray}
\label{F6superpot}
W^{Sp(4)}_{F=6} &=& \frac{1}{\Lambda^5_{Sp(4)}} \Big( \frac{1}{3} T_2 M_0^3 
                      + \frac{1}{2} M_0 M_1^2 \Big) \nonumber \\ 
W^{Sp(6)}_{F=6} &=& \frac{1}{\Lambda^7_{Sp(6)}} \Big( \frac{1}{3} T_2^2 M_0^3
                    + \frac{1}{2} T_3 M_1 M_0^2 - \frac{1}{2} T_2 M_0^2 M_2 +
                                          \nonumber \\
      & &            \frac{1}{4} M_0 M_2^2 + \frac{1}{4} M_1^2 M_2 \Big)
                                          \nonumber \\   
W^{Sp(8)}_{F=6} &=& \frac{1}{\Lambda^9_{Sp(8)}} \Big( \frac{1}{3} T_2^3 M_0^3
                    + \frac{1}{3} T_3^2 M_0^3 - \frac{1}{3} T_2 T_4 M_0^3
                    + T_2 T_3 M_0^2 M_1 + \nonumber \\
      & &             \frac{1}{2} T_4 M_0 M_1^2
                    + \frac{1}{2} T_4 M_0^2 M_2 + \frac{1}{2} T_2^2 M_0 M_1^2
                    - \frac{1}{2} T_2^2 M_2 M_0^2 +\frac{1}{6} T_3 M_1^3 -
                                          \nonumber \\
      & &             \frac{1}{2} T_3 M_0^2 M_3 - T_2 M_0 M_1 M_3 
                    - \frac{1}{2} T_2 M_1^2 M_2 + \frac{1}{4} M_0 M_3^2 +
                                          \nonumber \\  
      & &              \frac{1}{2} M_1 M_2 M_3 + 
                       \frac{1}{12} M_2^3 \Big),
\end{eqnarray}
where the $SU(6)$ flavor indices are contracted with $M^3\equiv
\epsilon^{ijklmn} M_{ij} M_{kl} M_{mn}$.
Note that the terms $T_2 M_0 M_1^2$ in $Sp(6)$ and $T_3 M_0 M_1 M_2$ and
$T_2 M_0 M_2^2$ in $Sp(8)$ are allowed by the symmetries but their
coefficients vanish.

We now present further non-trivial consistency checks which corroborate
the results for the superpotentials in Eq.~\ref{F6superpot}. First, we
add a mass term $m T_2$ for the antisymmetric tensor in the $Sp(4)$ theory.
This way we obtain a theory with six fundamentals and no antisymmetric tensor,
which is known to be in the confining phase with chiral symmetry
breaking~\cite{IntPoul}. 
The new superpotential of our theory is 
\begin{equation}
  W=T_2 (m+\frac{1}{\Lambda^5} M_0^3) + \frac{1}{\Lambda^5} M_0 M_1^2.
\end{equation}
The $T_2$ equation of motion $M_0^3=m \Lambda^5 = \tilde{\Lambda}^6$
forces non-zero vacuum expectation values for $M_0$. This vacuum
expectation value renders all components of $M_1$ massive via the
$M_0 M_1^2$ term in the superpotential. Furthermore, $T_2$ pairs up
to get a mass with the component of $M_0$ that lies in the direction
of the constraint $M_0^3=\tilde{\Lambda}^6$. Thus, we reproduce the results
of Ref.~\cite{IntPoul} for an $Sp(4)$ theory with six fundamentals
as required by consistency.

Next, consider breaking $Sp(4)$ to $SU(2)$ by giving vevs to $Q_5$ and
$Q_6$. In the ultraviolet theory two of the resulting $SU(2)$ doublets are
eaten by the Higgs mechanism. Six doublets remain, four from $Q_{1,2,3,4}$
and two from $A$. To see that our infrared theory reproduces
the correct result for an $SU(2)$ theory with six fundamentals, we
substitute the vevs into the $Sp(4)$ superpotential of
Eq.~\ref{F6superpot} to obtain the superpotential $W=\frac{1}{\Lambda^3}
{\rm Pf} M$. This is indeed the correct superpotential for the $SU(2)$
theory with six doublets if $M$ is identified with the $SU(2)$ meson matrix.

Another way of connecting our results to a known theory is to break $Sp(2 N)$
to $SU(2)^{N}$ by giving a vev of the following form to $A$:
\begin{equation}
\label{Avev}
 \langle A \rangle = i \sigma_2 v \otimes \left( 
            \begin{array}{cccc}
              \omega_1 & & & \\
               & \omega_2 & & \\
               & & \ddots & \\
               & & & \omega_{N}
            \end{array} \right),
\end{equation}
where $\omega_i$ are the $N$-th roots of 1. Note that this vev is 
traceless as required. The scales of the $SU(2)$ theories are related
to the scale of the $Sp(2N)$ theory by
$\Lambda_{Sp(2 N)}^{2 N+1} = \Lambda_{SU(2)_k}^3 v^{2 (N-1)} w_{-2 k}$.
Taking the limit $v\rightarrow \infty$ and $\Lambda_{Sp(2 N)}
\rightarrow \infty$ while holding the $\Lambda_{SU(2)}$'s fixed we obtain
a product of $N$ decoupled $SU(2)$ theories with six fundamentals each.
In the infrared theory we obtain the operator mapping from $Sp(2 N)$ to $SU(2)^{N}$  
by scaling out the large vevs from the $T$'s and $M$'s:
\begin{eqnarray}
\label{opmap}
  M_0 &\to& N_1 + N_2 + \ldots + N_{N} \nonumber \\
  M_1 &\to& (\omega_1 N_1 + \omega_2 N_2 + \ldots + \omega_{N} N_{N}) v
          \nonumber \\
  M_2 &\to& (\omega_2 N_1 + \omega_4 N_2 + \ldots + \omega_{2 N} N_{N}) v^2
          \nonumber \\
  \vdots & & \nonumber \\
  M_{N-1} &\to& (\omega_{N-1} N_1 + \omega_{2 (N-1)} N_2 + \ldots + 
                \omega_{N (N-1)} N_{N}) v^{N-1} \nonumber \\
  \langle T_2 \rangle &=& \langle T_3\rangle =\ldots =\langle T_{N-1} \rangle
              = 0 \nonumber \\
  \langle T_{N} \rangle &=& N v^{N},
\end{eqnarray}
where $N_i$ is the meson field of $SU(2)_i$. Substituting these expressions
together with the scale matching relations into the superpotentials of
Eq.~\ref{F6superpot}, we reproduce the correct superpotentials for the
$N$ decoupled $SU(2)$ theories
\begin{displaymath}
 W=\frac{1}{\Lambda^3_{SU(2)}} \left( N_1^3 + N_2^3 +\ldots + N_{N}^3
                               \right).
\end{displaymath}
We have checked this explicitly for $N=2$ and 4.

We now turn to the $F<6$ theories. These can be obtained by integrating
out flavors from the $F=6$ theory. To flow to the $F=4$ theories, we add a
mass term $m (M_0)_{56}$ to the superpotentials of Eq.~\ref{F6superpot}.
Taking the equation of motion with respect to $(M_0)_{56}$, we obtain
a quantum modified constraint for the fields of the $F=4$ theory.
More constraints follow from the equations
of motion of other massive mesons $(M_i)_{56}$. These constraints
are identical to the classical ones and ensure the equality of the
quantum and classical moduli spaces in the limit of large expectation values.
Thus, we find that the $Sp(2 N)$
with $F=4$ confines with chiral symmetry breaking. One of the classical
constraints is modified quantum mechanically whereas the other $N-1$
constraints remain unmodified. It is easy to check that the modified
constraint respects all symmetries of the theory, and that the $U(1)$ charges
do not allow a quantum modification of the others.
The large number of constraints reflects the much richer structure of
this theory compared to SUSY QCD.

There are several non-trivial consistency checks on this picture. As noted
above, for $F=4$ one can choose a $U(1)_R$ symmetry under which all the
superfields of the theory are neutral. Therefore, the $U(1)_R$ 
symmetry remains unbroken
even after chiral symmetry breaking. Since there are no non-zero R charges
around, no superpotential (which has R charge two) can be generated
dynamically. In the microscopic theory, both the global
$U(1)_R^3$ and $U(1)_R$ anomalies are equal to $1-6 N$.
In the macroscopic theory, these anomalies are exactly matched after
taking into account the $N$ constraints on the fields $M$ and $T$.
The maximal unbroken subgroup of the global symmetries is
$Sp(6) \times U(1)_R$. This leaves one further anomaly to check:
$Sp(6)^2 U(1)_R = -4 N$ in the microscopic and macroscopic theories.

We can enforce the constraints by using $N$ Lagrange multipliers
$\lambda_i$ in the superpotential:
\begin{eqnarray}
\label{F4superpot}
 W^{Sp(4)}_{F=4} &=& \lambda_1 \left( T_2 M_0^2 + \frac{1}{2} M_1^2 - 
                     \Lambda^6_{F=4}\right) + \lambda_2 M_0 M_1, \nonumber \\
 W^{Sp(6)}_{F=4} &=& \lambda_1 \left(T_2^2 M_0^2 + T_3 M_1 M_0 - T_2 M_0 M_2 +
                     \frac{1}{4} M_2^2 - \Lambda^8_{F=4} \right) + \nonumber \\
                 & & \lambda_2 \left(T_3 M_0^2 + M_1 M_2\right) + 
                     \lambda_3 \left(-M_0^2 T_2 + M_0 M_2 +\frac{1}{2} M_1^2
                                \right).
\end{eqnarray}

Having established the superpotential for the $F=4$ theories one can flow to
$F=2$ by adding another mass term $m (M_0)_{34}$. Integrating out the heavy
fields we obtain the superpotentials for $F=2$. We give only the $Sp(4)$
result:
\begin{equation}
\label{F2superpot}
  W^{Sp(4)}_{F=2} = \frac{\Lambda^7_{F=2} M_0}{2 T_2 M_0^2 - M_1^2}.
\end{equation}
This superpotential can also be obtained by breaking the $Sp(4)$ to $SU(2)
\times SU(2)$ with an appropriate vev of $A$. Instantons in the two $SU(2)$'s
give this superpotential. The superpotential for larger $N$'s are very
complicated, but they are straightforward to obtain by integrating out
flavors.

Finally, we can eliminate all fundamentals by adding one more mass term.
In the $Sp(4)$ case, we find two branches of the $F=0$
theory. One branch has a vanishing superpotential, while the other
has a superpotential generated by gaugino condensation:
\begin{equation}
\label{F0superpot}
  W=\pm \frac{\Lambda^4_{F=0}}{\sqrt{T_2}}.
\end{equation}
On the branch where 
$W=0$, the theory is in the confining phase without chiral symmetry
breaking. The only global symmetry of the microscopic theory is $U(1)_R$.
The $T_2$ operator matches all anomalies associated with this symmetry.
On the branch described by the superpotential of Eq.~\ref{F0superpot}
the theory has no stable vacuum.

The result for $F=0$ is not new. $Sp(4)$ and $SO(5)$ are isomorphic, and the
antisymmetric tensor of $Sp(4)$ corresponds to the fundamental of $SO(5)$.
An $SO(5)$ theory with one fundamental was described in~\cite{IntrSeiberg}
and the results are in complete agreement providing a very reassuring
consistency check on both the $Sp(2 N)$ as well as the $SO(N)$ analysis.

\section{Duality}

Based on the analogy with SUSY QCD, we expect there to be a dual description
for the theories with $F>6$. Finding simple duals with only one gauge group
for theories with tensor
representations has proven to be very difficult, no examples of such dualities
had been found prior to this work. After addition of a superpotential
that breaks chiral symmetries and simplifies the low-energy theories
by reducing the moduli space, several examples of dualities have been
given~\cite{Kutasov}. The $Sp(2 N)$ theory with an antisymmetric tensor
has been described in Ref.~\cite{ken}. Here, we tackle the theory without
a superpotential. We present a dual for $F=8$ and arbitrary $N$.
The dual is a generalization of the known $Sp(2)$ result~\cite{IntPoul},
where the antisymmetric tensor is simply a singlet. 

In our dual, the meson operators $M_k=Q A^k Q$ appear as fundamental fields, 
in addition
there is a dual gauge group $Sp(2 N)$ with an antisymmetric tensor,
eight dual quarks and a superpotential. The dual is defined by the following
field content
\begin{equation}
  \begin{array}{c|cccc}
      & Sp(2N)  & SU(8)   & U(1) & U(1)_R \\ \hline
    a & \Yasymm & 1       & -4   & 0 \\
    q & \Yfund  & \overline{\Yfund} & N-1 & \frac{1}{2} \\
   M_0&  1      & \Yasymm & 2N-2 & 1 \\
   M_1&  1      & \Yasymm & 2N-6 & 1 \\
   \vdots & \vdots & \vdots & \vdots & \vdots \\
   M_{N-1} & 1  & \Yasymm & -2N+2 & 1
  \end{array}
\end{equation}
and the superpotential 
\begin{equation}
\label{F8dualsup}
  W=q a^{N-1} q M_0 + q a^{N-2} q M_1 + \ldots + qq M_{N-1}.
\end{equation}
It is easy to see that the moduli spaces of the two theories agree.
In the electric theory, the flat directions are described by the
operators $T_k$ and $M_k$, while in the magnetic theory 
they are given by $t_k=a^k$ and $M_k$. 
The $\widetilde{M}_k=q a^k q$ directions are however
lifted by the $M_k$ equations of motion. 

A very non-trivial check on the duality is given by the `t~Hooft anomaly
matching conditions, which are all satisfied. Another important check
is to see that this dual flows to the $F=6$ theory after integrating out
two flavors in the electric theory. In the dual, this corresponds to
adding $m (M_0)_{78}$ to the superpotential of Eq.~\ref{F8dualsup}.
The equation of motion for $M_0$: $m=q a^{N-1} q$ forces non-zero vevs
for $a$ and $q$ which higgses the gauge group. The superpotential
terms give rise to masses for the $q$'s and the extra components of the $M$'s.
The superpotential of Eq.~\ref{F8dualsup} does not
reproduce the superpotential of Eq.~\ref{F6superpot}. The missing
terms are presumably generated by instantons in the broken $Sp(2 N)$.

As a final consistency check we consider breaking the electric theory
to $SU(2)^N$ by giving a vev to the $A$ field. In the limit of large vevs
this leads to $N$ decoupled $SU(2)$'s with eight doublets each. The magnetic
theory is also broken to $SU(2)^N$ by the corresponding vev of the $a$
field, again each $SU(2)$ group has eight doublets. After defining the
operator map in analogy with Eq.~\ref{opmap} and taking the vevs to infinity,
the superpotential of Eq.~\ref{F8dualsup} reduces to the correct
superpotential for $N$ factors of the dual of $SU(2)$.

We might also consider adding the operator ${\rm Tr} A^k$ to our theory
and its dual. We would expect to flow to the duals of the theories with
the superpotential described in Ref.~\cite{ken}. However, ${\rm Tr} A^k$
simply maps to ${\rm Tr} a^k$ in our dual, and our duals now look similar
to the duals in Ref.~\cite{ken}, but agree only for $k=N$. For other values
of $k$ even the sizes of the dual gauge groups are different. It would
be interesting to understand the connection between our duals and those
of Ref.~\cite{ken}.

Our dual seems to contradict the expectation that the dual
of a theory with arbitrary superpotential can be obtained from the
dual of the theory with no superpotential.
It would be very interesting to know whether this duality can be extended
for $F>8$, or if the case $F=8$ where the electric and magnetic theories
have identical gauge degrees of freedom (``self-dual'') is special.

Another type of duals can be found for arbitrary values of $F$ by using
the deconfinement method of Ref.~\cite{Berkooz}. Applying the 
``deconfinement modulus'' of Ref.~\cite{Martin} and choosing the
additional global symmetry to be $SU(2)$ one obtains the first dual. This
dual is an $Sp(F-4)\times Sp(2 N-2)$ theory, in which the $Sp(F-4)$ gauge
group has only fundamentals, while the $Sp(2 N-2)$ group contains an
antisymmetric tensor and $2 F-4$ fundamentals. Iterating this procedure
$N-1$ times, we arrive at the gauge group $Sp(F-4)\times Sp(2 F-8) \times
\ldots \times Sp(N(F-4)-2)$. All the gauge groups in this chain, except
for the $Sp(N(F-4)-2)$, contain an antisymmetric tensor. The $Sp(N(F-4)-2)$
group has only fundamentals and is very strongly coupled. 
Unfortunately, this dual
description is of limited use since there is no value
of $F$ for which all gauge groups are weakly coupled.

\section{Applications to Model Building}
Now, we will use the exact results obtained in Section 2 for building
models of dynamical supersymmetry breaking and of SUSY models of
compositeness.

\subsection*{Models of Dynamical Supersymmetry Breaking}
Dine, Nelson, Nir and Shirman recently noted
that one can often find new models of dynamical supersymmetry breaking
by reducing the gauge group of a known model~\cite{Dine}. Several
examples of models obtained by using this method were presented in
Refs.~\cite{Dine,us}.

We now show that one can obtain a model of dynamical supersymmetry breaking
by reducing the $SU(5)$ model with a ${\bf 10}$ and a ${\bf \bar{5}}$ to
$Sp(4) \times U(1)$. The field content of the $Sp(4) \times U(1)$ theory
is given by
\begin{equation}
  \begin{array}{c|cc}
     & Sp(4)    & U(1) \\ \hline
   A & \Yasymm  & 2    \\
  Q_1& \Yfund   & -3   \\
  Q_2& \Yfund   & -1   \\
  S_1& 1        & 2    \\
  S_2& 1        & 4   
 \end{array}
\end{equation}
We add the tree-level superpotential
\begin{equation}
\label{wtree}
 W_{tree}= Q_1 Q_2 S_2 + Q_1 A Q_2 S_1.
\end{equation}
The $Sp(4)$ invariants and their $U(1)$ charges are
\begin{equation}
  \begin{array}{c|c}
          & U(1) \\ \hline
   T_2=A^2& 4    \\
   M_0=Q_1 Q_2 & -4 \\
   M_1=Q_1 A Q_2 & -2 \\
   S_1           & 2 \\
   S_2           & 4
  \end{array}
\end{equation}
The $S_1$ and $S_2$ equations of motion set $M_0$ and $M_1$ to zero,
and the $U(1)$ charges of the remaining $Sp(4)$ invariants are all of the
same sign. Therefore, all classical flat directions are lifted by the
superpotential of Eq.~\ref{wtree}. 

The full superpotential also contains a term generated by the $Sp(4)$
dynamics, as described in Eq.~\ref{F2superpot}:
\begin{equation}
  W=Q_1 Q_2 S_2 + Q_1 A Q_2 S_1 + \frac{\Lambda^7 (Q_1 Q_2)}{2 (A^2)
                                        (Q_1 Q_2)^2 -(Q_1 A Q_2)^2}.
\end{equation}
The tree-level superpotential preserves a non-anomalous $U(1)_R$ symmetry,
which has to be broken because of the presence of the dynamically-generated
term. Therefore, one expects supersymmetry to be dynamically broken,
and indeed the equations of motion are contradictory.

One can also built models of dynamical supersymmetry breaking by making
use of quantum modified moduli spaces~\cite{IntThomas}. Consider for example
an $Sp(4)$ theory  with $F=4$ and additional singlets $S_i$ coupled via the
following superpotential
\begin{equation}
  W_{tree}=S_1 M_0 + S_2 M_1 + S_3 T_2.
\end{equation}
The fields $S_1$ and $S_2$ transform as antisymmetric tensors under the $SU(4)$
flavor symmetry. The equations of motion for the singlet fields set $M_0=
M_1=T_2=0$. Such a solution does not lie on the quantum modified moduli space
described by Eq.~\ref{F4superpot}, therefore supersymmetry is dynamically
broken in this theory.

\subsection*{Composite Model Building}

These new results may have interesting applications for SUSY models of
compositeness~\cite{compost,effsusy}.
For example, it is an intriguing possibility that the three families of
the standard model are composite and arise after confinement of a
preonic $Sp(6)$ theory with an antisymmetric tensor and $F=6$ fundamentals.
As discussed above, the infrared spectrum of this theory contains the
three meson fields $M_0=QQ$, $M_1=QAQ$,
and $M_2=QA^2Q$ which transform as antisymmetric tensors of
the flavor symmetry. Thus, if an $SU(5)$ subgroup of the $SU(6)$ flavor
symmetry was weakly gauged, one would find three generations of $\bf 10$'s
of $SU(5)$ after the $Sp$ confinement. Alternatively, one might just gauge
an $SU(3) \times SU(2) \times U(1)$ subgroup to obtain the supersymmetric
standard model.

To illustrate this idea we present a simple toy model that reproduces
the particle content of a SUSY $SU(5)$ GUT
plus two vector like pairs of ${\bf 5} + {\bf \bar{5}}$. The model consists
of the fields given in the following table:
\begin{equation}
 \begin{array}{c|c|cc}
    & Sp(6) & SU(5) & SU(6) \\ \hline
  A & \Yasymm   & 1     & 1 \\
 Q& \Yfund    & \Yfund& 1 \\
 Q'& \Yfund & 1 & 1 \\
 \overline{Q} & 1 & \overline{\Yfund} & \Yfund
 \end{array}
\end{equation}
After confinement of the $SP(6)$ we obtain
\begin{equation}
 \begin{array}{c|cc}
  & SU(5) & SU(6) \\ \hline
  M_0,\, M_1,\, M_2 & \Yasymm   & 1  \\
 H_0,\, H_1,\, H_2& \Yfund    & 1 \\
 T_2,\, T_3 & 1 & 1 \\
 \overline{Q} & \overline{\Yfund} & \Yfund
 \end{array}
\end{equation}
and the non-perturbatively generated superpotential
\begin{eqnarray}
 W &=& H_0 ( \frac{T_2^2}{\Lambda^2} M_0^2 - \frac{T_3}{\Lambda} M_0 M_1 +  
 \frac{T_2}{\Lambda} M_0 M_2 -M_2^2) + \nonumber\\
   & & H_1 (-\frac{T_3}{\Lambda} M_0^2+M_1 M_2) + H_2(\frac{T_2}{\Lambda}  
 M_0^2-M_0 M_2+M_1 M_1).
\end{eqnarray}
When the composite field $H_0$ is identified with $H_U$ of the supersymmetric
standard model this results in the following Yukawa coupling matrix for
the up-type quarks is obtained:
\begin{equation}
Y_{U} \sim \left(
 \matrix{
   {T_2}^2/\Lambda^2 & T_3/\Lambda & T_2/\Lambda \cr
   T_3/\Lambda & 0 & 0\cr
   T_2/\Lambda & 0 & 1}\right)
\end{equation}
Assume that the vacuum expectation values of the composite $T_i$ are
somewhat smaller than the scale $\Lambda$ where the $SP(6)$ confines, the
theory generates a hierarchical structure of Yukawa couplings dynamically.
Such vevs for the $T_i$'s could be enforced by adding a tree-level
superpotential.

This way the hierarchical structure of Yukawa
couplings that is so difficult to obtain in conventional models arises
automatically as a consequence of the strong preon dynamics!
From a low-energy point of view the origin of the hierarchical
structure can be understood in terms of a ``horizontal" $U(1)$ symmetry under
which the generations transform with different charges.
In the ultraviolet this ``horizontal" symmetry is simply the anomaly-free
$U(1)$. The difference in charges between the generations
reflects the fact that the $M_i$ contain different numbers of the preon
field $A$.
 
Down quark and lepton Yukawa couplings arise from higher-dimensional
operators that have to be present before the confinement of the $Sp$
gauge group. Requiring that the superpotential respects the above-mentioned
$U(1)$ symmetry, one finds down Yukawa couplings with a
somewhat smaller hierarchy than in the up sector. This is nice because 
the quark mass ratios in the down sector are smaller than in the up sector.
Obviously, this toy model is too simple to account for all details
of the standard model. This is left for a future investigation.

\section{Conclusions}

We have investigated the non-perturbative behavior of SUSY $Sp(2N)$
gauge theories containing an antisymmetric tensor. The theory with $F=6$
confines without chiral symmetry breaking. The low-energy degrees of freedom
are the mesons $M_k=Q A^k Q$ and $T_k= {\rm Tr} A^k$. They interact via
a confining superpotential, which was determined exactly by demanding
consistency with the classical constraints. Starting from the $F=6$
theory we obtain non-trivial results for the theory with less flavors:

-- The $F=4$ theory confines with chiral symmetry breaking; an interesting
result is that there are $N$ constraints, one of which is modified
quantum mechanically.

-- The $F=2$ theory has an instanton-generated superpotential and a runaway
vacuum.

-- The $F=0$ theories have two disconnected branches of vacua: one with
no superpotential and a moduli space of vacua, and one with a runaway
vacuum.

We have constructed a dual description to the $F=8$ theory, which is
the first of this kind in the literature. It is not obvious how this dual
relates to the known duals of theories with added superpotential.
It would be very interesting to understand
whether this dual for $F=8$ is due to a fortunate coincidence or it can
be generalized to larger values of $F$.

We also presented two interesting applications of our results to model
building. One application is to construct models of dynamical supersymmetry
breaking. Another is for building composite models. One can embed the
supersymmetric standard model into an
$Sp(6)\times SU(5)$ gauge theory, where $Sp(6)$ has an antisymmetric tensor
and $F=6$. Confinement in $Sp(6)$ results in composite quarks and leptons.
A mass hierarchy for the standard model particles arises naturally as a
result of the confining superpotential of $Sp(6)$.

\section*{Acknowledgments}
We are grateful to Lisa Randall for discussions and for comments on the
manuscript.
M.S. is supported in part by the U.S.\ Department of Energy under grant
\#DE-FG02-91ER40676. C.C. and W.S. are supported in part by the U.S.\
Department of Energy under cooperative agreement \#DE-FC02-94ER40818.
M.S. would like to thank the Aspen Center for Physics and the Institute
for Nuclear Theory at the University of Washington for their hospitality.

While this manuscript was in preparation we became aware of preprint
hep-th/9607200 by P.~Cho and P.~Kraus~\cite{scoop} who independently
obtained our results of Section 2.


\begin{thebibliography}{99}
 \bibitem{exact}  N.~Seiberg, \prd{49} (1994) 6857 
 \bibitem{duality} N.~Seiberg, \npb{435} (1995) 129
 \bibitem{ADS} I.~Affleck, M.~Dine and N.~Seiberg, \npb{256} (1985) 557
 \bibitem{IntrSeiberg} K.~Intriligator and N.~Seiberg, \npb{444} (1995) 125 
 \bibitem{IntPoul} K.~Intriligator and P.~Pouliot, \pl{B335} (1995) 471
 \bibitem{PoppitzTriv}  E.~Poppitz and S.P.~Trivedi, \pl{B365} (1996) 125
 \bibitem{Pouliot}  P.~Pouliot, \pl{B367} (1996) 151 
 \bibitem{Kutasov} D.~Kutasov, \pl{B351} (1995) 230; \\
                   D.~Kutasov and A.~Schwimmer, \pl{B354} (1995) 315; \\
                   R.G.~Leigh and M.J.~Strassler, \pl{B356} (1995) 492; \\
                   D.~Kutasov, A.~Schwimmer and N.~Seiberg,
                            \npb{459} (1996) 455 
 \bibitem{Martin} M.A.~Luty, M.~Schmaltz and J.~Terning, hep-th/9603034
 \bibitem{ILS} K.~Intriligator, R.G.~Leigh and N.~Seiberg, \prd{50} (1994) 1092
 \bibitem{PST} E.~Poppitz, Y.~Shadmi and S.P.~Trivedi, hep-th/9605113, \\ 
               hep-th/9606184
  \bibitem{ken} K.~Intriligator, \npb{448} (1995) 187
 \bibitem{Witten} E.~Witten, \pl{117B} (1982) 324
 \bibitem{Berkooz} M.~Berkooz, \npb{452} (1995) 513
 \bibitem{Dine} M.~Dine, A.N.~Nelson, Y.~Nir and Y.~Shirman,
                \prd{53} (1996) 2658 
 \bibitem{us} C.~Cs\'aki, L.~Randall and W.~Skiba, hep-th/9605108; \\   
              C.~Cs\'aki, L.~Randall, W.~Skiba and R.G.~Leigh, hep-th/9607021
 \bibitem{IntThomas} K.~Intriligator and S.~Thomas, hep-th/9603158
 \bibitem{compost} A.E.~Nelson and M.J.~Strassler, hep-ph/9607362; \\
    M.J.~Strassler, \pl{B367} (1996) 119
 \bibitem{effsusy} A.G. Cohen, D.B. Kaplan, and A.E. Nelson, hep-ph/9607394 
 \bibitem{scoop} P.~Cho and P.~Kraus, hep-th/9607200
\end{thebibliography}
\end{document}